\documentclass[prl,twocolumn,
tightenlines,superscriptaddress,nofootinbib,showpacs]{revtex4}
\usepackage{amssymb,latexsym}
\usepackage{amsmath,amsbsy,bbm}
\usepackage{epsfig,bm}
\usepackage{graphicx,comment}
\usepackage{color}
\DeclareMathOperator{\tr}{tr}

\begin{document}
\def\a{{\alpha}}
\def\b{{\beta}}
\def\d{{\delta}}
\def\D{{\Delta}}
\def\e{{\varepsilon}}
\def\g{{\gamma}}
\def\G{{\Gamma}}
\def\k{{\kappa}}
\def\l{{\lambda}}
\def\L{{\Lambda}}
\def\m{{\mu}}
\def\n{{\nu}}
\def\o{{\omega}}
\def\O{{\Omega}}
\def\S{{\Sigma}}
\def\s{{\sigma}}
\def\th{{\theta}}
\newcommand{\mnod}{\stackrel{\circ}{M}}

\def\ol#1{{\overline{#1}}}

\def\Dslash{\ol D\hskip-0.65em /}
\def\Dslashe{D\hskip-0.65em /}
\def\Pslash{\ol P\hskip-0.65em /}
\def\lslash{l\hskip-0.35em /}
\def\Pslashe{P\hskip-0.65em /}

\def\Dtslash{\tilde{D} \hskip-0.65em /}

\def\CPT{{$\chi$PT}}
\def\QCPT{{Q$\chi$PT}}
\def\PQCPT{{PQ$\chi$PT}}
\def\tr{\text{tr}}
\def\str{\text{str}}
\def\diag{\text{diag}}
\def\order{{\mathcal O}}

\def\cC{{\mathcal C}}
\def\cB{{\mathcal B}}
\def\cT{{\mathcal T}}
\def\cQ{{\mathcal Q}}
\def\cL{{\mathcal L}}
\def\cO{{\mathcal O}}
\def\cA{{\mathcal A}}
\def\cQ{{\mathcal Q}}
\def\cR{{\mathcal R}}
\def\cH{{\mathcal H}}
\def\cW{{\mathcal W}}
\def\cM{{\mathcal M}}
\def\cD{{\mathcal D}}
\def\cN{{\mathcal N}}
\def\cP{{\mathcal P}}
\def\cK{{\mathcal K}}
\def\Qt{{\tilde{Q}}}
\def\Dt{{\tilde{D}}}
\def\St{{\tilde{\Sigma}}}
\def\cBt{{\tilde{\mathcal{B}}}}
\def\cDt{{\tilde{\mathcal{D}}}}
\def\cTt{{\tilde{\mathcal{T}}}}
\def\cMt{{\tilde{\mathcal{M}}}}
\def\At{{\tilde{A}}}
\def\cNt{{\tilde{\mathcal{N}}}}
\def\cOt{{\tilde{\mathcal{O}}}}
\def\cPt{{\tilde{\mathcal{P}}}}
\def\cI{{\mathcal{I}}}
\def\cJ{{\mathcal{J}}}
\def\cb{{\cal B}}
\def\cbb{{\overline{\cal B}}}
\def\ct{{\cal T}}
\def\ctt{{\overline{\cal T}}}

\def\eqref#1{{(\ref{#1})}}

 
\title{Chiral Corrections and the Axial Charge of the Delta}

\author{Fu-Jiun~Jiang}
\email[]{fjjiang@itp.unibe.ch}
\affiliation{Institute for Theoretical Physics, Bern University, Sidlerstrasse 5, CH-3012 Bern, Switzerland}

\author{Brian~C.~Tiburzi}
\email[]{bctiburz@umd.edu}
\affiliation{Maryland Center for Fundamental Physics, Department of Physics, University of Maryland, College Park, MD 20742-4111, USA}

\date{\today}


\begin{abstract}
Chiral corrections to the delta axial charge
are determined using heavy baryon chiral perturbation theory. 
Knowledge of this axial coupling is necessary to assess
virtual-delta contributions to nucleon and delta observables.
We give isospin relations useful for a lattice determination of the axial coupling.
Furthermore we detail partially quenched chiral
corrections, which are relevant to address partial
quenching and/or mixed action errors in lattice calculations
of the delta axial charge.
\end{abstract}

\maketitle

{\bf Introduction}.---
Numerical simulations of QCD
on spacetime lattices provide a first principles method to study 
non-perturbative regime of QCD~\cite{DeGrand:2006aa}.
For the last decade, lattice QCD has made dramatic progress   
due to enlarged computing
resources, and advances in numerical algorithms. 
Even with considerable progress, 
lattice simulations are still restricted to unphysically large quark masses, 
and lattice sizes that are not much larger than typical hadronic length scales.
Fortunately, low-energy hadronic physics is 
dominated by pion interactions which can be studied systematically using chiral
perturbation theory ($\chi$PT). 
Today lattice data in conjunction with $\chi$PT enable first principles
predictions, and in turn, the investigation 
of the low-energy effective theory.

Properties of baryons can be addressed systematically using 
$\chi$PT~\cite{Jenkins:1991jv,Jenkins:1991es}. 
There are, however, notable complications to this proposal. 
The number of \emph{a priori} unknown low-energy constants for baryons 
is larger at next-to-leading order compared to those of mesons. 
The chiral expansion in powers of 
$m_\pi^2 / \Lambda_\chi^2$, 
where 
$m_\pi$ 
is the pion mass and 
$\Lambda_\chi$
the chiral symmetry breaking scale, 
is accompanied by an expansion in 
$m_\pi / M_N$, 
where 
$M_N$ 
is the nucleon mass in the chiral limit. 
Finally the nearby delta-resonances
can lead to important contributions.
For a recent review of delta physics, see~\cite{Pascalutsa:2006up}.  
The axial charge of the nucleon, for example, receives 
important virtual contributions from pion-delta intermediate states
because the  delta-nucleon mass splitting is about the same 
size as the pion mass, and the axial couplings
$G_{\D N}$, 
and 
$G_{\D\D}$ 
are of order one.
We focus on the axial charge of the delta, 
$G_{\D \D}$. 
The value of this parameter is largely unknown for two reasons. 
Firstly and obviously, the short mean lifetime of the delta complicates
experimental extraction of this coupling.
Secondly the value of 
$G_{\D \D}$
has been inferred from one-loop chiral computations
of various baryon observables. 
These computations are incomplete, however, 
as they only roughly estimate or completely neglect 
local contributions from unknown low-energy constants.
There are too many unknowns to extract reliable information about 
$G_{\D\D}$
from chiral computations alone. 
Lattice QCD simulations 
can remedy this.

Early work on the delta using lattice QCD
centered on electromagnetic
moments and delta-to-nucleon electromagnetic transitions%
~\cite{Leinweber:1992hy,Leinweber:1992pv}.
These calculations have been refined recently~\cite{Alexandrou:2004xn,Alexandrou:2007dt,Alexandrou:2007we}, 
by including the effects of dynamical quarks,
and reaching much lower pion masses  
$\sim 350 \, \texttt{MeV}$. 
The axial nucleon-to-delta transition has also been 
studied on the lattice for the first time~\cite{Alexandrou:2006mc,Alexandrou:2007zz}. 
The delta axial charge, $G_{\D\D}$, can also be 
determined using lattice QCD. 
A necessarily component for this study 
is the stability of the delta. 
Above the decay threshold, 
$m_\pi > \D$, where $\D$ is the
delta-nucleon mass splitting in the chiral limit, the
delta will be a stable particle on the lattice. 
Its static properties can be calculated
and the effective theory is used, in turn, 
to extrapolate down to the physical point,
where the axial matrix element becomes 
complex valued.
Lighter pion masses too can be employed. 
The reason being that the delta decays
via $p$-wave pion emission, and the
available momentum modes 
are restrictive enough to keep 
the delta \emph{lattice stabilized}. 
Lattice study of the delta at lighter pion
masses requires more care, especially
with volume effects~\cite{Bernard:2007cm},
but could better control the chiral expansion
of delta properties.

In this work, we
relate various delta matrix elements 
to the axial coupling 
$G_{\D\D}$. 
The connected $\D^{++}$ matrix element 
provides a convenient starting point
for lattice simulations. 
We compute the one-loop 
chiral corrections to the 
delta axial charge, 
and investigate its pion 
mass dependence.
The modulus $|G_{\D \D}|$
is shown to be relatively stable
with respect to chiral corrections.
Finally we perform
the partially quenched 
chiral computation.

{\bf Delta Axial Matrix Elements}.--- 
%
%
%
There are various axial current matrix elements in the quartet of $\Delta$-resonances. 
Several of these can be used at zero momentum transfer to define the axial charge, 
$G_{\D\D}$. 
The remaining choices are then completely determined as a product of
$G_{\D\D}$
and isospin Clebsch-Gordan coefficients. 
To arrive at the conventional definition, 
we define the axial charge through the relation
\begin{equation} \label{eq:AxialDef}
\langle \D^{++} | J^{3}_{\mu 5} | \D^{++} \rangle
-
\langle \D^{-} | J^{3}_{\mu 5} | \D^{-} \rangle
= 
G_{\D\D}  
\mathcal{M}_\mu
,\end{equation}
in which appears the axial current
$J_{\mu 5}^a
= 
\ol Q  \gamma_\mu \gamma_5  T^a Q$
, 
where the isospin generators 
are given by 
$T^a = \frac{1}{2} \tau^a$, 
with 
$\tau^a$ 
as Pauli matrices.
The factor 
$\mathcal{M}_\mu$
encodes the spin structure of the
forward matrix element, namely
$\mathcal{M}_\mu
= 
\ol U {}^\nu(P) \gamma_\mu \gamma_5 U_\nu(P)
$, 
where 
$U_\nu(P)$ 
is a Rarita-Schwinger spinor.
The definition in Eq.~\eqref{eq:AxialDef}
can be easily utilized to calculate 
$G_{\D\D}$
on the lattice.
Because differences of isosinglet matrix
elements vanish, 
we can write Eq.~\eqref{eq:AxialDef} as
\begin{equation} \label{eq:Lattice}
\langle \D^{++} | \ol u \gamma_\mu \gamma_5 u | \D^{++} \rangle_{\text{conn.}}
= G_{\D\D} \mathcal{M}_\mu 
,\end{equation}
where the subscript denotes only connected quark contractions.

For completeness, we specify the other matrix elements
from which the axial charge can be deduced. 
The isospin changing matrix elements are related by the 
Wigner-Eckart theorem. We consider only 
$\D I = +1$ 
for ease, and find
\begin{eqnarray}
\langle \D^{++} | J^+_{\mu 5} | \D^+ \rangle
&=& 
\frac{1}{\sqrt{3}} C \mathcal{M}_\mu
\notag \\
\langle \D^{+} | J^+_{\mu 5} | \D^0 \rangle
&=& 
\frac{2}{3} C \mathcal{M}_\mu
\notag \\
\langle \D^{0} | J^+_{\mu 5} | \D^- \rangle
&=& 
\frac{1}{\sqrt{3}} C \mathcal{M}_\mu
\label{eq:Changing}
,\end{eqnarray}
where 
$C$ 
is proportional to the common reduced matrix element.
Using isospin, one can relate the isospin transitions
to differences of 
$\D I = 0$ 
matrix elements and thus leads to
\begin{eqnarray}
\frac{1}{\sqrt{3}} 
\langle \D^{++} | J^+_{\mu 5} | \D^+ \rangle
&=&
\langle \D^{++} | J^{3}_{\mu 5} | \D^{++} \rangle
-
\langle \D^{+} | J^{3}_{\mu 5} | \D^{+} \rangle
\notag \\
\frac{1}{2} \langle \D^{+} | J^+_{\mu 5} | \D^0 \rangle
&=& 
\langle \D^{+} | J^{3}_{\mu 5} | \D^{+} \rangle
-
\langle \D^{0} | J^{3}_{\mu 5} | \D^{0} \rangle
\notag  \\
\frac{1}{\sqrt{3}}
\langle \D^{0} | J^+_{\mu 5} | \D^- \rangle
&=& 
\langle \D^{0} | J^{3}_{\mu 5} | \D^{0} \rangle
-
\langle \D^{-} | J^{3}_{\mu 5} | \D^{-} \rangle .\nonumber \\
\label{eq:Differences}
\end{eqnarray}
Combining all three of these relations shows
that 
$C = G_{\D\D}
$. 
 One can then utilize the 
$\D I = 1$ 
relations in 
Eq.~\eqref{eq:Changing} to determine the axial 
charge, or the matrix element differences in 
Eq.~\eqref{eq:Differences}. The former 
are directly tied to weak interaction phenomenology,
while the latter are straightforward to implement on 
the lattice, e.g.,~Eq.~\eqref{eq:Lattice}. 
As with the other baryon axial couplings, there are no
disconnected diagrams to evaluate.
When we generalize to partially quenched theories (or additionally 
mixed lattice actions), the above relations between matrix elements 
remain valid because of the vector isospin symmetry in the valence sector.

{\bf Chiral Computations}.---
%
The 
$SU(2)_L \otimes SU(2)_R$ 
symmetry of two-flavor QCD
is spontaneously broken down to the vector subgroup. 
The low-energy dynamics are described
by pseudo-Goldstone pions
emerging from spontaneous chiral symmetry 
breaking. 
These modes, 
$\phi$,
are non-linearly realized 
in the coset field 
$\Sigma \equiv \xi^2 = \exp ( 2 i \phi / f )$,
where 
\begin{equation}
\phi 
= 
\begin{pmatrix}
\pi^0 / \sqrt{2}  & \pi^+ \\
\pi^- & - \pi^0 / \sqrt{2}
\end{pmatrix}
,\end{equation}
and $f = 132 \, \texttt{MeV}$
is the pion decay constant.
Pion dynamics are described 
at leading order%
\footnote{Here we adopt the standard power counting: 
$\partial \sim m_\pi \sim \varepsilon$, where $\varepsilon$ is a small
parameter.}
 by the effective Lagrangian
\begin{equation}
\cL 
= 
\frac{f^2}{8}
\tr (\partial^\mu \Sigma^\dagger \partial_\mu \Sigma)
+ 
\frac{f^2 m_\pi^2}{8}
\tr ( \Sigma^\dagger + \Sigma)
\label{eq:ChPT}
.\end{equation}

\begin{figure}
\begin{center}
\includegraphics[width=0.35\textwidth]{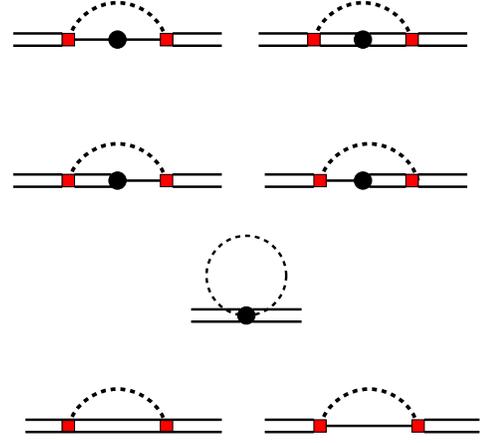}
\end{center}
\caption{One-loop diagrams which contribute 
non-analytic terms to the axial charge of the delta. 
Mesons are represented by a dashed line, while the single 
and double lines are the symbols for nucleons and deltas respectively. 
The solid circle is an insertion of the axial current operator.
The wave function renormalization diagrams are shown 
at the bottom. 
}
\label{fig0}
\end{figure}

\begin{figure}
\begin{center}
\includegraphics[width=0.35\textwidth]{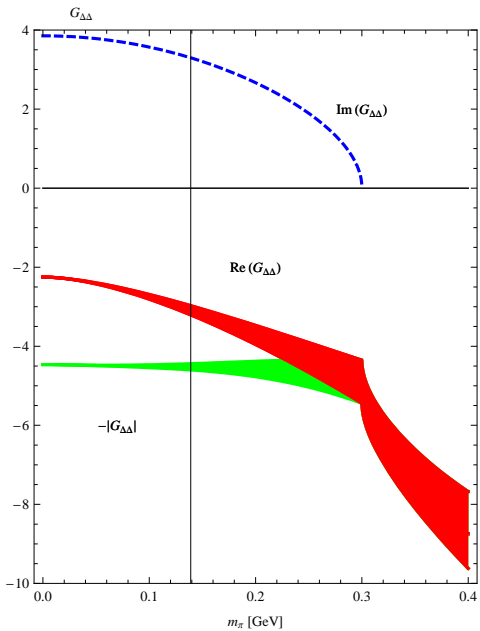}
\end{center}
\caption{
Dependence of the delta axial charge $G_{\D\D}$ on the pion mass. 
The plotted bands account for uncertainty arising from the unknown NLO coupling. }
\label{figChiral}
\end{figure}

The baryons are contained in $SU(2)_V$ multiplets: 
a doublet $N$ of spin-$1/2$ nucleons and a quartet
$T$ of spin-$3/2$ deltas. The Lagrangian up to NLO
describing the nucleons, deltas and their interactions 
with pions is given by
\begin{eqnarray}
\cL 
&=& 
i \ol N v \cdot D N
- 
i \ol T_\mu v \cdot D T^\mu
+ 
\D \ol T_\mu T^\mu
+
2 g_A 
\ol N S \cdot A N
\notag \\
&+&
g_{\D N}
\left( \ol T_\mu A^\mu N + \ol N A_\mu T^\mu \right)
+ 
2 g_{\D\D}
\ol T_\mu S \cdot A T^\mu
\label{eq:L}
,\end{eqnarray}
where $v_\mu$ is the baryon velocity, 
and $S_\mu$ the spin operator, see
~\cite{Jenkins:1991jv,Jenkins:1991es}
for further details.
The leading order axial current
derived from Eq.~\eqref{eq:L}
produces the result $G_{\D \D} = g_{\D \D}$. 
Beyond this order there is a local contribution
from the NLO current (which only differs
from the LO current by an insertion of the quark mass)
as well as loop contributions depicted in Fig.~\ref{fig0}.
Evaluating these contributions, we find the delta axial charge
\begin{eqnarray}
G_{\Delta\Delta} 
&=& 
g_{\Delta\Delta}Z_{\Delta} 
-
\frac{1}{(4 \pi f)^2} 
\Bigg[
2 g_{\Delta\Delta} 
{\cal L}(m_\pi,\mu)
\Bigg( 1 + \frac{121}{324} g_{\D \D}^2 \Bigg)
\nonumber \\
&+& 
g^2_{\Delta N} 
\Bigg(
\frac{8}{9} 
g_{\Delta\Delta}
{\cal K}(m_{\pi},-\Delta,\mu)
- 
g_A   {\cal J}(m_{\pi},-\Delta,\mu) 
\Bigg) \Bigg]
\notag \\
&+&
A m^2_{\pi}  
\, .
\label{matrixiso}
\end{eqnarray}
%
%
%
The non-analytic functions appearing above, 
namely, 
${\cal L}(m,\mu)$, 
${\cal J}(m,\D,\mu)$, 
and
${\cal K}(m,\D,\mu)$ 
are given in~\cite{Jiang:2008aq}.
The delta wavefunction renormalization 
$Z_\D$ 
appears in \cite{Tiburzi:2004rh}.
Lastly the constant $A$ is the 
parameter appearing in the NLO 
delta axial current.
In Fig.~\ref{figChiral}, we plot the 
pion mass dependence of the 
axial charge $G_{\D\D}$. 
We fix the tree-level axial couplings 
at their $SU(4)$ values, and renormalize
the loop graphs so that
${\text{Re}}(G_{\D\D}) = g_{\D\D}$ in the chiral limit.
To qualitatively understand the contributions from 
the NLO coupling $A$, we vary the renormalization 
scale $\mu$ from $800$ to $1200 \, \texttt{MeV}$ 
and further plot the real and imaginary parts of $G_{\D\D}$
along with (minus) the modulus.
The figure shows that the modulus is 
governed by perturbative chiral corrections, 
not the real or imaginary parts.

{\bf Partially Quenched Chiral Computations}.---
%
%
The partially quenched generalization of the Lagrangian in Eq.~\eqref{eq:ChPT}
reads~\cite{Sharpe:2001fh}
\begin{eqnarray} \label{eq:Llead}
\cL =  
\frac{f^2}{8}
\str \Big(\partial^{\mu}\Sigma^\dagger \partial_\mu\Sigma\Big)
+ 
\frac{\lambda f^2}{4} \str\Big[m_Q (\Sigma^\dagger + \Sigma)\Big]
-
\mu_0^2  \Phi_0^2
,\notag\\ \end{eqnarray}
where the mass matrix has been generalized to
\begin{equation}
m_Q = \diag (m, m, m_j, m_j, m, m)
\label{massmatrix2}
,\end{equation}
in the isospin limit of the valence and sea sectors.
The valence pion mass $m_\pi$ is given by
$m_\pi^2 = 2 \lambda m$, while the sea pion
mass $m_{jj}^2 = 2 \lambda m_j$. 
Finally the valence-sea pion mass $m_{ju}^2$
is given by the average of the two.\footnote{%
These statements may be modified in the case
of a mixed lattice action. It is straighforward
to take this into account given partially 
quenched expressions for observables, 
see~\cite{Bar:2005tu,Tiburzi:2005is,Chen:2007ug}.
For example, the valence-sea meson 
mass receives an additive renormalization
because no symmetry relates the valence and sea sectors.
}
The flavor singlet field $\Phi_0$ appearing in 
Eq.~\eqref{eq:Llead} is $\Phi_0 = \str \Sigma / \sqrt{2}$
and has been retained as a device.
The axial anomaly allows the singlet 
mass parameter $\mu_0$ to be on the order
of the chiral symmetry breaking scale. 
Subsequently integrating out the singlet yields
the correct interactions in the flavor neutral 
sector of the theory. These are described
by the so-called hairpin vertex, see~\cite{Sharpe:2001fh}.

Baryons fields are described in terms
of the $\bf{70}$-dimensional super-multiplet 
$\cB_{ijk}$ containing spin-$1/2$ baryons,
and 
the $\bf{44}$-dimensional supermultiplet
$\cT^{\mu}_{ijk}$ containing the spin-$3/2$ baryons.
For the embedding of the familiar nucleons
and delta into these supermultiplets, as well as their free and interaction 
Lagrangian, 
see~\cite{Labrenz:1996jy,Savage:2001dy,Chen:2001yi,Beane:2002vq}.  
At leading order, 
the PQ$\chi$PT delta axial current is given by \cite{Beane:2002vq}
\begin{eqnarray}
J_{\mu5}^{a}
& = & 
2{\cal H}\Big(\overline{\cal T}^\nu S_\mu {\overline{\tau}{}^{a}_{\xi +}}{\cal T}_\nu\Big)
,\label{eq:LOaxialcurrent}
\end{eqnarray}
where 
$\overline{\tau}{}^a_{\xi +} = {1\over 2}(
\xi\overline{\tau}{}^a\xi^\dagger
+\xi^\dagger\overline{\tau}{}^a\xi)$,
and 
$\ol{\tau}^a$ 
are partially quenched
extensions of the isospin generators.
Because the axial charge 
$G_{\D \D}$
can be determined from connected
quark contractions,~Eq.~\eqref{eq:Lattice},
we follow~\cite{Tiburzi:2005hg} and choose the upper 
$2\times2$ 
block of 
$\overline{\tau}^a$ 
to be the ordinary isospin generators
and zeros elsewhere. 
The constant $\cH$ can be determined
from matching, i.e.~$\cH = g_{\D \D}$.

\begin{figure}
\begin{center}
\includegraphics[width=0.35\textwidth]{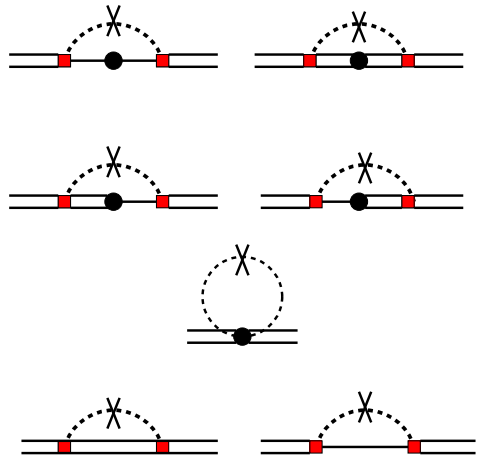}
\end{center}
\caption{Additional one-loop diagrams contributing to the delta axial charge in \PQCPT.  
The cross on the loop meson is a hairpin contribution.
}
\label{fig0PQ}
\end{figure}

Since we work to next-to-leading order (NLO) in the chiral expansion, 
we further require contributions from the NLO
axial current. These involve one insertion of the quark 
mass matrix $m_{Q}$
\begin{widetext}
\begin{eqnarray}
\d J_{\mu5}^{a}&=& \frac{16 \lambda}{f^2}\Bigg[
t_1 
\ctt^{kji}_{\mu}\{ \overline{\tau}^a_{\xi +}\,,\, 
{\cal M}_+\}^n_i\ S_\mu \ct_{njk}^{\mu}
+
t_2 (-)^{\eta_l(\eta_j+\eta_n)}
\ctt^{kji}_{\mu}(\overline{\tau}^a_{\xi +})^l_i 
( {\cal M}_+)^n_j S_\mu \ct_{lnk}^{\mu} 
\nonumber \\
&+&
t_3 \ctt^{kji}_{\mu} (\overline{\tau}^a_{\xi +})^l_i S_\mu \ct_{ljk}^{\mu}
\ {\rm str}( {\cal M}_+) 
+
t_4 \ctt^{kji}_{\mu} S_\mu \ct_{ijk}^{\mu} 
\ {\rm str}(\overline{\tau}^a_{\xi +} {\cal M}_+ )\Bigg]\,,
\label{eq:PQNLO}
\end{eqnarray}
where  
the mass operator ${\mathcal M}_{+}$ is defined by:
${\mathcal M}_+ = \frac{1}{2}\left(\xi^\dagger m_Q \xi^\dagger + \xi m_Q \xi\right)$.
When working to tree-level, there are only two independent contributions from
the NLO current: one is proportional to the valence pion mass squared and the 
other is proportional to the sea pion mass squared. Compared to $SU(2)$ 
$\chi$PT there is thus one additional parameter to be determined.
At NLO, there are additionally non-analytic contributions
arising from the loop diagrams in Figs.~\ref{fig0} and \ref{fig0PQ}. 
Evaluating these contributions, we find the partially quenched 
delta axial charge
\begin{eqnarray}
G_{\Delta\Delta} 
&=& 
g_{\Delta\Delta}Z_{\Delta} 
-
\frac{1}{(4 \pi f)^2} 
\Bigg\{
2 g_{\Delta\Delta} 
{\cal L}(m_{ju},\mu)
+
\frac{11}{9}
g^3_{\Delta\Delta}
\Bigg[
\frac{2}{3} {\cal L}(m_{\pi},\mu)
+
\frac{4}{9} {\cal L}(m_{ju},\mu)
+ {\cal R}(\eta_{u},\eta_{u},\mu)
\Bigg]
\nonumber \\
&+& 
g^2_{\Delta N} 
\Bigg[ 
\frac{8}{9} 
g_{\Delta\Delta}
{\cal K}(m_{ju},-\Delta,\mu)
+ 
(g_A + g_1)  {\cal J}(m_{\pi},-\Delta,\mu) 
-( 2 g_A + g_1 )   {\cal J}(m_{ju},-\Delta,\mu) 
\Bigg]
\Bigg\}
+ 
\cA m^2_{\pi}  
+ 
\cB m_{jj}^2  . \quad  
\label{matrixisoPQ}
\end{eqnarray}
\end{widetext}
The partially quenched 
wavefunction renormalization 
$Z_\D$ 
appears in \cite{Tiburzi:2004rh},
and the function $\mathcal{R}(\phi, \phi',\mu)$
which arises from hairpins is given in~\cite{Jiang:2008aq}.
The constants $\cA$ and $\cB$ are shorthands for linear
combinations of coefficients from the NLO partially quenched 
delta axial current, Eq.~\eqref{eq:PQNLO}. 
With Eq.~\eqref{matrixisoPQ}, 
one can extrapolate $G_{\D\D}$ 
in both valence and sea quark masses.
Finally with trivial modifications, 
the mixed action extrapolation can be performed.

\smallskip
\smallskip

\begin{acknowledgments}
F.-J.J. wished to thank C. W. Kao for discussions. This work is 
supported in part by the U.S.\ Dept.~of Energy, Grant 
No.DE-FG02-93ER-40762 (B.C.T.) and by the Schweizerischer 
Nationalfonds (F.-J.J.). 
\end{acknowledgments}


%

\bibliography{hb}

\end{document}